\documentclass[prd,aps,oamsmath,amssymb,nofootinbib,preprintnumbers]
{revtex4}
\usepackage{graphicx}
\usepackage{dcolumn}
\usepackage{bm}
\usepackage{amsmath}
\usepackage{amsthm}
\usepackage{multirow}
\usepackage{enumerate}

\usepackage{amsfonts}
\usepackage{euscript,bbm}
\usepackage{ifthen}
\usepackage{psfrag}
\usepackage{slashed}
\usepackage{hyperref}
\usepackage{gensymb}
\usepackage[utf8]{inputenc}
\usepackage{color}

\def\ls{\mathrel{\lower4pt\vbox{\lineskip=0pt\baselineskip=0pt
           \hbox{$<$}\hbox{$\sim$}}}}
\def\gs{\mathrel{\lower4pt\vbox{\lineskip=0pt\baselineskip=0pt
           \hbox{$>$}\hbox{$\sim$}}}}
\def\drawbox#1#2{\hrule height#2pt

\hbox{\vrule width#2pt height#1pt \kern#1pt
              \vrule width#2pt}
              \hrule height#2pt}

\def\Asym#1#2{\vcenter{\vbox{\drawbox{#1}{#2}
              \kern-#2pt       
              \drawbox{#1}{#2}}}}

\def\nn{\nonumber}

\newcommand{\be}{\begin{equation}}
\newcommand{\ee}{\end{equation}}
\newcommand{\bea}{\begin{eqnarray}}
\newcommand{\eea}{\end{eqnarray}}
\providecommand{\e}[1]{\ensuremath{\times 10^{#1}}}

\newcommand{\gsim}{\lower.7ex\hbox{$\;\stackrel{\textstyle>}{\sim}\;$}}
\newcommand{\lsim}{\lower.7ex\hbox{$\;\stackrel{\textstyle<}{\sim}\;$}}

\newcommand{\met} {{E\!\!\!\!/_{T}}}

\begin{document}

\begin{flushright}
MI-TH-1546
\end{flushright}

\title{Interpretation of the diphoton excess at CMS and ATLAS}

\author{Bhaskar Dutta$^{1}$}
\author{Yu Gao$^{1}$}
\author{Tathagata Ghosh$^{1}$}
\author{Ilia Gogoladze$^{2}$}
\author{Tianjun Li$^{3,4}$}

\affiliation{$^{1}$ Mitchell Institute for Fundamental Physics and Astronomy, \\
Department of Physics and Astronomy, Texas A\&M University, College Station, TX 77843-4242, USA \\
$^{2}$ Bartol Research Institute, Department of Physics and Astronomy, \\
University of Delaware, Newark, DE 19716, USA \\
$^{3}$ State Key Laboratory of Theoretical Physics and Kavli Institute for
Theoretical Physics China (KITPC), \\ Institute of Theoretical Physics,
Chinese Academy of Sciences, Beijing 100190, P. R. China \\
$^{4}$ School of Physical Electronics,  University of Electronic Science
and Technology of China, Chengdu 610054, P. R. China}

\begin{abstract}
We consider the diphoton resonance at the 13 TeV LHC in a consistent model with new scalars 
and vector-like fermions added to the Standard Model (SM), which can be constructed from
orbifold grand unified theories and string models. The gauge coupling unification can be achieved, neutrino masses 
can be generated radiatively, and electroweak vacuum stability problem can be solved. 
To explain the diphoton resonance, we study a spin-0 particle, and discuss various associated final states. We also constrain the couplings and number of the introduced heavy multiplets for the new resonance's width at 5 or 40 GeV.
\end{abstract}

\maketitle


\section{Introduction}

The recent 13 TeV CMS~\cite{bib:CMS_diphoton} and ALTAS~\cite{bib:ATLAS_diphoton} runs have reported a narrow two-photon resonance with
an invariant mass near 750 GeV, at a combined 3$\sigma$ level of credence. Combined with fluctuations
from previous 8 TeV data, the excess is reported around 3$\sigma$ at CMS~\cite{bib:LHCseminar} and 4$\sigma$ at ATLAS~\cite{bib:LHCseminar}.

The narrow diphoton resonance at 750 GeV, if confirmed by future LHC updates, will strongly indicate
a massive non-Standard Model (SM) spin-even state. A spin-1 state does not decay into two photons
due to the Landau-Yang theorem. An interesting possibility is that the new state $X$ being SM gauge singlet
but couples to the SM particles at loop-level via heavy new-physics scalars and vector-like fermions that
are charged under SM gauge groups. As an extension to the SM fermionic sector, we assume a heavy (TeV scale)
generation of both quarks and leptons, denoted as $Q$ and $L$ respectively. Their vector-like couplings
avoid anomaly. It is also interesting to understand the implications of these states for dark matter,
neutrino masses, grand unification etc. Such kind of models can be realized in the orbifold
Grand Unified Theories (GUT).

The next section~\ref{sect:model} we discuss the extensions to the SM and the loop-induced effective couplings.
In Section~\ref{sect:signal} we present the benchmark  parameter range of the model to explain the diphoton excess.
In Section~\ref{sect:associated} we discuss the imminent potential tests of associated collider signals.
We further discuss the possibilities of grand unification, neutrino masses, and dark matter, etc,
in Section~\ref{sect:theory_discussion}, and then conclude in Section ~\ref{sect:conclusion}.

\section{An economical SM extension}
\label{sect:model}

To accommodate for the diphoton signal, we can classify the spins and parities of
 the resonance particles as $0^+$, $0^-$, and $2^+$, since the vector particle can be
excluded due to the Landau-Yang theorem.

\begin{table}[!b]
\begin{tabular}{c|c|c|c|c|c|c|c}
\hline
 & $\kappa_1$ & $\kappa_2$ & $\kappa_3$ & & $\kappa_1$ & $\kappa_2$ & $\kappa_3$ \\
 \hline
~($Q$, ${\bar Q}$)~ & ~$\frac{\lambda_Q g_Y^2}{96\pi^2 M_Q}$~ &  ~$\frac{3 \lambda_Q g_2^2}{32\pi^2 M_Q}$~ &
 ~$\frac{\lambda_Q g_3^2}{8\pi^2 M_Q}$~ &
~($L$, ${\bar L}$)~ & ~$\frac{\lambda_L g_Y^2}{32\pi^2 M_L}$~ &  ~$\frac{ \lambda_L g_2^2}{32\pi^2 M_L}$~ &
 ~0~\\
\hline
~($U$, ${\bar U}$)~ & ~$\frac{\lambda_U g_Y^2}{12\pi^2 M_U}$~ &  0 &
 ~$\frac{\lambda_U g_3^2}{16\pi^2 M_U}$~ &
~($E$, ${\bar E}$)~ & ~$\frac{\lambda_E g_Y^2}{16\pi^2 M_E}$~ &  0 &
~0~\\
\hline
~($D$, ${\bar D}$)~ & ~$\frac{\lambda_D g_Y^2}{48\pi^2 M_D}$~ &  0 &
 ~$\frac{\lambda_D g_3^2}{16\pi^2 M_D}$~ &
 &  &   &
 ~~\\
\hline
\end{tabular}
\caption{The coefficient of $\kappa_i$s ($i=1,2,3$) for different vector-like particles. Please note the effective couplings $\kappa_i$s can obtained from the above coefficients by multiplying them by the loop function $A_{1/2}(\tau_F)$ ($F=Q,U,D,L,E$), presented in Eq.~\ref{loop_function_1}}
\label{T1Ki}
\end{table}

First, let us consider the CP-even scalar particle $X$ with mass around 750 GeV, similar to that of Standard Model Higgs~\cite{Carmi:2012in, Bonne:2012im, Aguilar-Saavedra:2013qpa},
to generate the couplings between $X$ and SM gauge fields, we introduce the vector-like particles
 $F$ and ${\bar F}$.
For simplicity, we only consider the vector-like particles whose quantum numbers are the same as
the SM fermions. The relevant Lagrangian is
\be
{\cal L}_{BSM} = \lambda_F X \bar{F} F +  \frac{M_X^2}{2} |X|^2 + M_F \bar{F} F +  \text{ kinetic terms}.
\label{eq:L_int}
\ee
where $F=Q, U, D, L$, and $E$.

\begin{figure}[!t]
\includegraphics[scale=0.5]{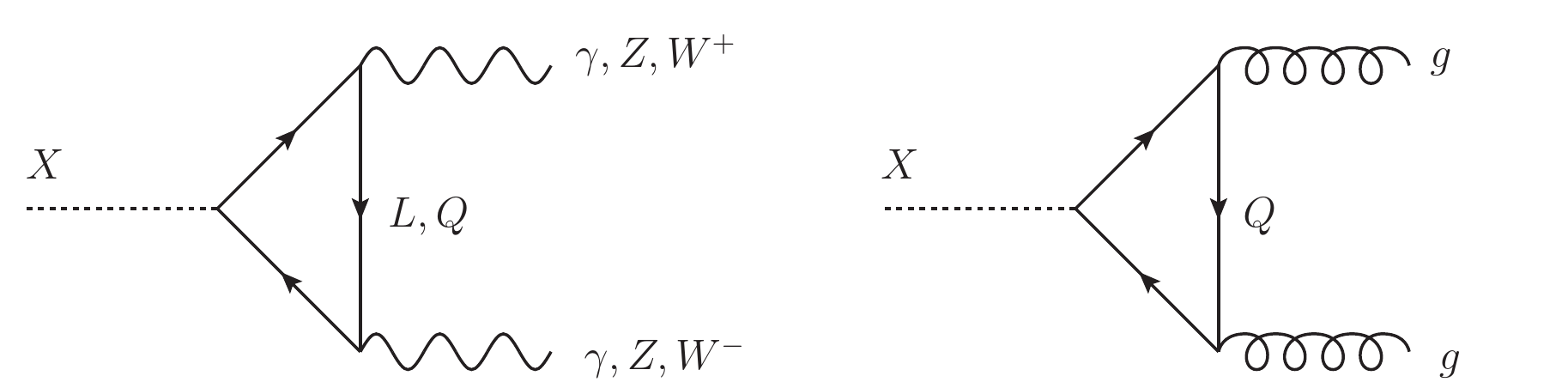}
\caption{ Heavy fermion loops that couple $s$ to a pair of SM gauge bosons}
\label{fig:feynman_loop}
\end{figure}

For heavy $F$ masses, an effective $XVV$-type vertex can be induced by $F$ triangle loops in Fig.~\ref{fig:feynman_loop}, as
\begin{eqnarray}
\label{Scalar}
\mathcal{L}_{eff.} =  \kappa_1 X B_{\mu \nu} B^{\mu \nu} + \kappa_2 X W^{j}_{\mu \nu} W^{j \mu \nu} +  \kappa_3 X G^{a}_{\mu \nu} G^{a \mu \nu}~,~
\end{eqnarray}
where $B_{\mu \nu}$, $W^{j}_{\mu \nu}$ and $G^{a}_{\mu \nu}$ represents the field strength tensor of the SM gauge
bosons of the $U(1)_Y$, $SU(2)_L$ and $SU(3)_c$ groups, respectively, with $j=1,2,3$ and $a=1,2,..8$ are the indices
of the adjoint representations of $SU(2)_L$ and $SU(3)_c$ respectively. We present the $\kappa_i$ for different $F$ and ${\bar F}$ in Table~\ref{T1Ki}.

One can also consider the CP-odd scalar particle $X$ with mass around 750 GeV.
The relevant Lagrangian is
\be
{\cal L}_{BSM} = \lambda_F X \bar{F} i \gamma_5 F +  \frac{M_X^2}{2} |X|^2 + M_F \bar{F} F +  \text{kinetic terms},
\label{eq:L_inta}
\ee

where the $\kappa_i$ for different $F$ and ${\bar F}$ are similar to those in Table~\ref{T1Ki}. The CP-odd and even cases should give identical diphoton signal rate (although the spin correlation in final state kinematics may differ) and we will restrict to the formalism for the CP-even case after this point. { Also, we will use $L,Q$ to denote vector-like new leptons and quarks collectively for collider signal discussions.}

The effective couplings of Eq.~\ref{Scalar}, After rotation to the physical gauge boson states, can be written as,
\bea
 \kappa_{\gamma \gamma } & = & \kappa_1 \cos^{2} \theta_W + \kappa_2 \sin^{2} \theta_W \, , \nn \\  
 \kappa_{Z Z } & = &  \kappa_2 \cos^{2} \theta_W + \kappa_1 \sin^{2} \theta_W \, , \nn \\
 \kappa_{Z \gamma } & = &  (\kappa_2 - \kappa_1) \sin 2 \theta_W \, , \nn \\
 \kappa_{WW } & = & 2 \kappa_2 \, , \nn\\
 \kappa_{gg } & = & \kappa_3 \, .      
\label{eq:eff_couplings_singlets}
\eea 

where $\theta_W$ is the Weinberg mixing angle. As mentioned before the effective couplings $\kappa_i$s can be obtained from the coefficients presented in Table~\ref{T1Ki} by multiplying them by the loop function, $A_{1/2}(\tau_F)$ (where $F=Q,U,D,L,E$) with a spin-$1/2$ particle in the loop. The loop function $A_{1/2}(\tau)$ with $\tau = 4 M^{2}/M^{2}_X$  is given by, 
\bea
A_{1/2}(\tau) &=& 2 \tau [1+ (1-\tau) f(\tau)], 
\label{loop_function_1}
\eea

with

\bea
f(x)= 
	\begin{cases}
		arcsin^2[1/\sqrt{x}],  & \mbox{if } x \geq 1 \\
		-\dfrac{1}{4}[ln \dfrac{1+\sqrt{1-x}}{1-\sqrt{1-x}} - i \pi]^2, & \mbox{if } x < 1.
	\end{cases}
\label{loop_function_2}
\eea


With the minimal extension of Eq.~\ref{eq:L_int}, the lightest neutral component of $L,Q$ (for instance the heavy `neutrino' in an isodoublet $L$), if present, is stable and can be a dark matter candidate. In case the lightest component is charged, it can pair up with a SM fermion into a stable compound state.
 Alternatively, a small mixing between the heavy $L,Q$ and the SM fermions may be introduced via Yukawa type interaction,
\bea
y\bar{Q}q_{R}^{SM}H \hspace{1cm} & (\text{for isodoublet } Q)  \nn \\
y\bar{Q}^{SM}Q H \hspace{0.9cm}  & (\text{for isosinglet} Q)
\eea
which then allows the heavy $Q$ to decay into their SM fermionic counterparts plus a SM boson. Here we use $Q^{SM}, q_R^{SM}$ to denote SM doublet and singlet quarks and $H$ for the Higgs doublet. $Q$ can be strongly pair-produced at the LHC and leads to long-lived ionizing heavy particle (if stable) or a two 2jet+2$h$/2$V$ (if unstable) final state. A massive $M_Q >1$ TeV can be consistent with these bounds (see~\cite{ATLAS-VLQ} and references therein). Similar mixings can also be introduced for $L$.  However, since $L$ is only weakly produced, a $\sim 400$ GeV scale mass is most likely out of the current LHC reach.

\begin{figure}[!t]
\includegraphics[scale=0.5]{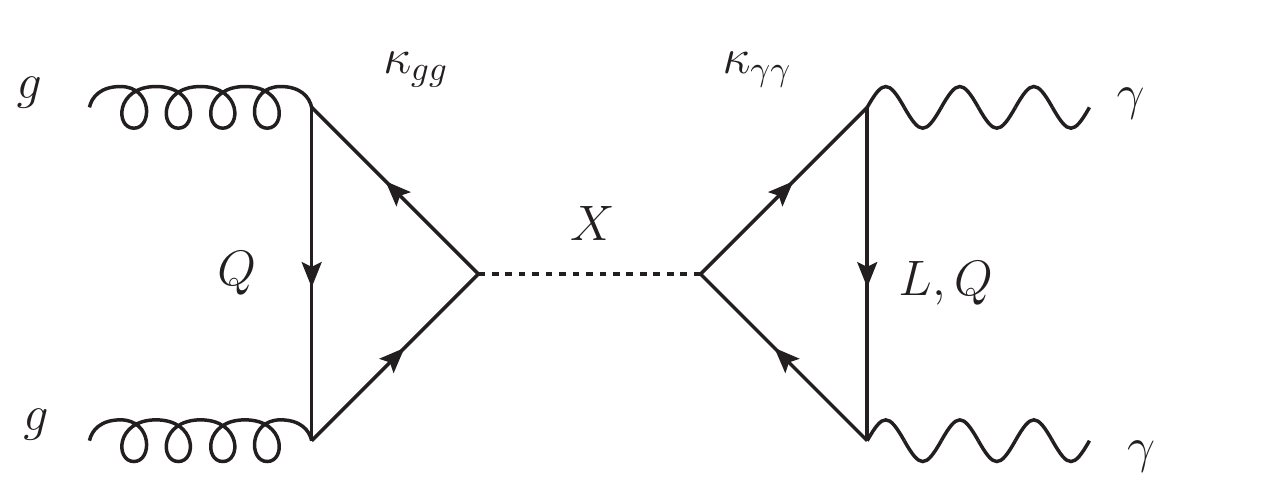}
\caption{Loop-level production and diphoton decay of a singlet heavy scalar $s$. }
\label{fig:feynman_proc}
\end{figure}

\section{The diphoton signal}
\label{sect:signal}

Two of the couplings in Eq.~\ref{eq:eff_couplings_singlets}, $\kappa_{gg}$ and $\kappa_{\gamma\gamma}$, can be responsible for LHC diphoton process as shown in Fig.~\ref{fig:feynman_proc}. For decay width up to a few percent of the mass, the cross-section can be given in the narrow-width approximation as
\bea
   \sigma_{\gamma\gamma} & = & \dfrac{\pi^2}{8}\dfrac{\Gamma(X\rightarrow gg)}{M_X}\times \text{BR}(X \rightarrow \gamma\gamma) \times \Big [\dfrac{1}{s}\dfrac{\partial \mathcal{L}_{gg}}{\partial \tau} \Big ], \nn \\   
\dfrac{\partial \mathcal{L}_{gg}}{\partial \tau} & = & \int \limits_{0} dx_1 dx_2  f_{g}(x_1) f_g(X_2) \delta(x_1 x_2 - \frac{M^{2}_X}{s}),
\eea
where $\sqrt{s}=13$ TeV and $f_g$ denotes the gluon parton distribution function inside a proton, with $x$ being the fraction of each beam's energy carried away by the corresponding gluon.

One should note that the experimentally observed width of the resonance is appreciably large.
ATLAS suggested a width as large as $\Gamma_X = 6\% M_X$. However, the data collected so far is inconclusive. Ref.~\cite{Falkowski} has performed a likelihood analysis to fit both CMS and ATLAS data and
checked for their compatibility against the 8 TeV data as well. 
They have found that for the combined run-I and run-II data, a width of 5 GeV provides almost as good a fit as a width of 40 GeV. Hence, in this analysis we will present benchmark points (BP) with $\Gamma_X=5$ GeV as well as $\Gamma_X=40$ GeV. For $\Gamma_X=5$ GeV Ref.~\cite{Falkowski} has found the best fit $\sigma_{\gamma\gamma} = 2.4$ fb for $M_X=750$ GeV. However, a $\sigma_{\gamma\gamma} \sim 0.5 -4.5$ fb can satisfy the resonance at $95 \%$ CL. In contrast, for $\Gamma_X=40$ GeV the best fit is obtained for $M_X = 730$ GeV with $\sigma_{\gamma\gamma} = 6$ fb. The corresponding $95 \%$ CL range is $\sim 2-10$ fb. If we fix $M_X$ at 750 GeV the best fit $\Gamma_X$ and $\sigma_{\gamma\gamma}$ are 30 GeV and 4.8 fb, respectively.

This resonant cross-section has the parameter dependence
\be
\sigma_{\gamma\gamma} \propto  \frac{\kappa_{g g}^2\kappa_{\gamma\gamma}^2}{\Gamma_X} 
 \propto  \frac{\kappa_{g g}^2\kappa_{\gamma\gamma}^2}{8 \kappa_{g g}^2 + \kappa_{\gamma\gamma}^2}, 
\label{eq:kappa_dependence}
\ee 
where $\Gamma_X$ is the total decay width of the resonance. 
 As is evident from Eq. ~\ref{eq:kappa_dependence}, $\kappa_{gg}$ can be uniquely determined by the experimentally measured $\sigma_{\gamma\gamma}$ in two cases: 
\begin{enumerate}[(I)]
\item only vector-like quarks ($Q,U,D$) are present in the model and $\kappa_{\gamma\gamma} \propto \kappa_{gg}$,
\item when $\kappa_{gg}\ll \kappa_{\gamma\gamma}$ where the total width is dominated by large $L$ or $E$ loop contributions. 
\end{enumerate} 
 
 \subsection{Case I: vector-like quark only scenarios}

\begin{figure}[!t]
\includegraphics[scale=0.25]{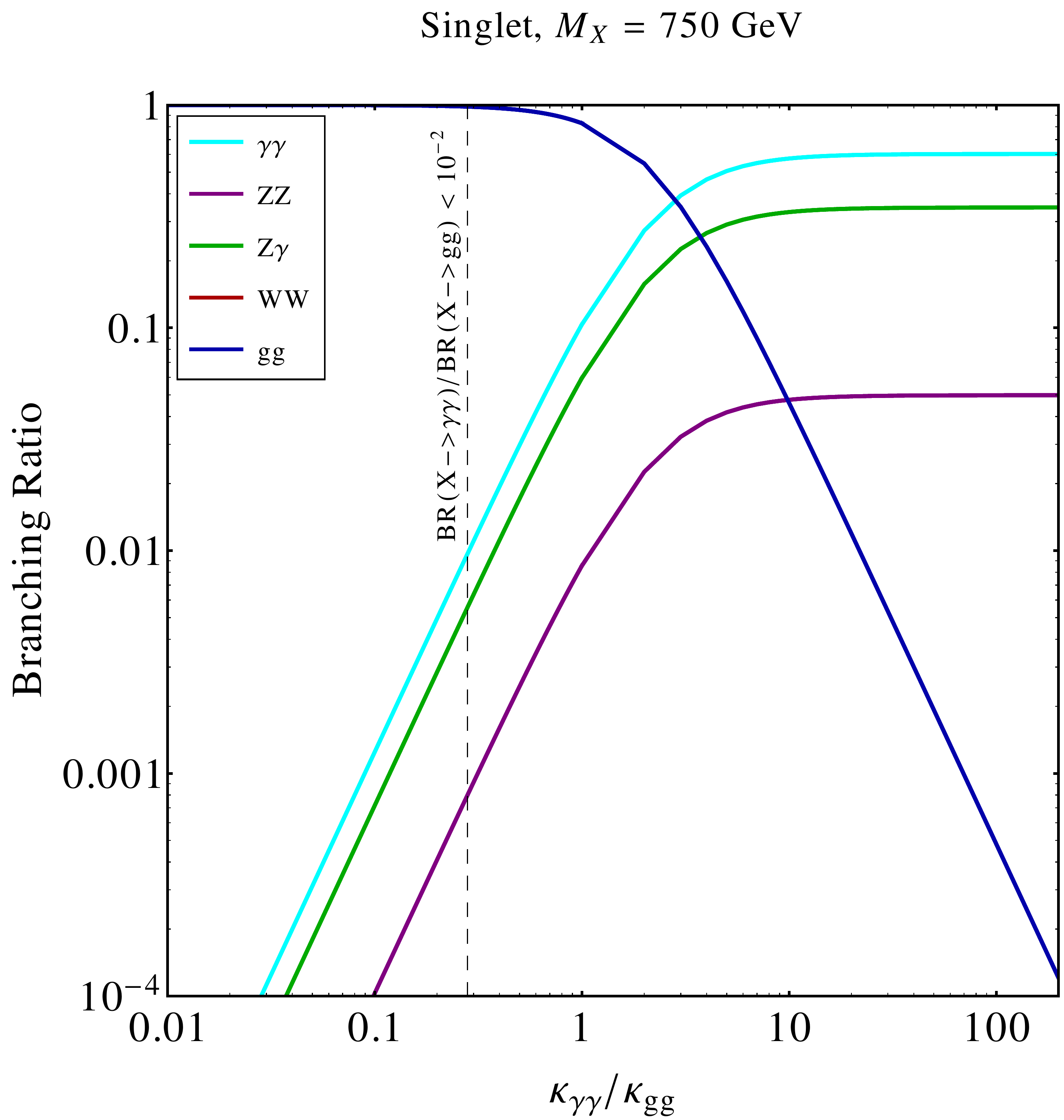}
\includegraphics[scale=0.25]{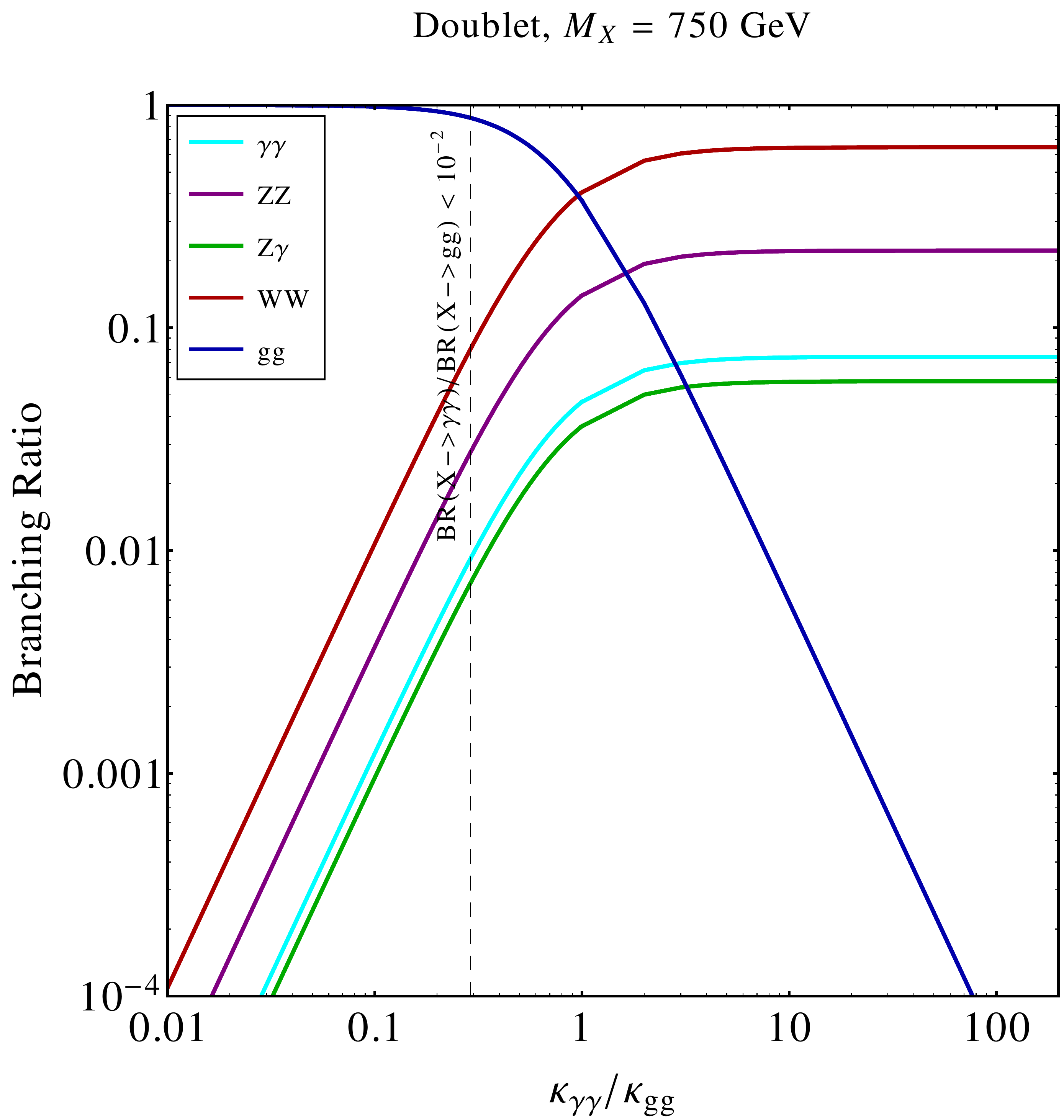}
\caption{The 750 GeV scalar decay branchings versus the relative ratio between its coupling to $L$ and $Q$. The left panel shows the case that $L,Q$ are isosinglets and the right panel shows the isodoublet case.}
\label{fig:branchings}
\end{figure}

In the $Q$-only scenario, the correlation between $\kappa_{gg}$ and $\kappa_{\gamma\gamma}$ is fixed: $\frac{\kappa_{\gamma\gamma}}{\kappa_{gg}} =0.05$ when $Q$ is an isodoublet, 0.09 when $Q$ is an up-type isosinglet, and 0.02 when a down-type isosinglet. For $M_X=750$ GeV we found that 
\be 
\kappa_{gg} =\left\{
\begin{array}{l}
8.7 \times 10^{-5} \hspace{0.1 cm} \text{GeV}^{-1}, \hspace{1cm} \text{ doublet}\\
5.4 \times 10^{-5} \hspace{0.1 cm} \text{GeV}^{-1}, \hspace{1cm} \text{u-type isosinglet}\\
2.1 \times 10^{-4} \hspace{0.1 cm} \text{GeV}^{-1}, \hspace{1cm} \text{d-type isosinglet}\\
\end{array}
\right.
\ee
gives the best-fit signal rate of $\sigma_{\gamma\gamma} = 4.8$ fb~\cite{Falkowski}. The coupling needed for the $d$-type singlet case is larger due to its small $\kappa_{\gamma\gamma}$ due to reduced electric charge. 

In Table~\ref{tab:Q_BP_no_inv} we present two sets of BPs for doublet, u-type isosinglet and d-type isosinglet cases. For the first set, we fix the width to $\Gamma_X = 5$ GeV and evaluate the associated couplings and cross-sections in various diboson channels. For the second set of BPs we determine the couplings by fixing $\sigma_{\gamma\gamma} = 4.8$ fb and perform the same calculations. We have set $M_Q = 1$ TeV for both set of calculations. One can notice from Table~\ref{tab:Q_BP_no_inv} that we need $N_Q\lambda_Q \sim 3 - 17$ to fit either $\Gamma_X = 5$ GeV or $\sigma_{\gamma\gamma} = 4.8$ fb due to small $\kappa_{\gamma\gamma}$ in the $Q$-only scenarios. Hence,  multiple numbers of $Q$ fields are then predicted to keep the coupling $\lambda_Q$ perturbative. We also show the BR for decay of $X$ in various diboson channels in Fig.~\ref{fig:branchings}. Another alarming prediction from $Q$-only scenario is the very small $BR(X\rightarrow \gamma\gamma)< 10^{-3}$ (esp. for down-type $Q$ due to its small electric charge), as shown in Fig.~\ref{fig:branchings}. Since $\sigma_{gg} =\sigma_{\gamma\gamma}\frac{BR(X\rightarrow g g)}{BR(X\rightarrow \gamma\gamma)}$, and $BR(X\rightarrow g g)\sim 1$, a very tiny $BR(X\rightarrow \gamma\gamma)< 10^{-2}$ may boost $\sigma_{gg}$ above the current dijet bound at 2 pb at 8 TeV~\cite{CMS:2015neg}. We discuss each BP, along with possible constraint from the aforementioned CMS dijet bound with more detail in the following paragraph.

\begin{table}[!b]
\setlength\tabcolsep{5pt}
\begin{tabular}{|c|c c c|c c c|}
\hline\hline
 Type & doublet & u-type singlet & d-type singlet & doublet & u-type singlet & d-type singlet \\
 \hline
 $Q$-only     & BP-1 & BP-2 & BP-3 & BP-4 & BP-5 & BP-6 \\
 \hline     
$M_X$ [GeV] & \multicolumn{3}{c|}{750} & \multicolumn{3}{c|}{750}  \\
$\Gamma_X$ [GeV] & \multicolumn{3}{c|}{5} & 2.09 & 0.79 & 12.7 \\
$N_Q \lambda_Q$ & 5.18 &  10.4 & 10.4 & 3.35 & 4.16 & 16.6 \\
\hline
$\kappa_{\gamma \gamma}$ [GeV$^{-1}$] & 7.41$ \e{-6}$ & 1.19$ \e{-5}$ & 2.99$ \e{-6}$ &  \multicolumn{3}{c|}{4.76$ \e{-6}$} \\
$\kappa_{gg}$ [GeV$^{-1}$] & \multicolumn{3}{c|}{1.36$ \e{-4}$} & 8.75$ \e{-5}$ & 5.44$ \e{-5}$ & 2.17$ \e{-4}$  \\
$\kappa_{WW}$ [GeV$^{-1}$] & 5.77$ \e{-5}$ & 0 & 0 & 3.73$ \e{-5}$ & 0 & 0 \\
\hline
$\sigma_{\gamma \gamma}$ [fb] & 11.9 & 31.4 & 1.97 & \multicolumn{3}{c|}{4.8}  \\
$\sigma_{Z Z}$ [fb] & 99.6 & 2.59 & 0.16 & 41.6 & 0.41 & 0.41 \\
$\sigma_{Z \gamma}$ [fb] & 57.4 & 18.1 & 1.13 & 24 &  2.87 & 2.87  \\
$\sigma_{W W}$ [fb] & 337 & 0 & 0 & 141.1 & 0  &  0 \\
$\sigma_{g g}$ [fb] & 3.18$ \e{4}$ & 3.27$ \e{4}$ & 3.28$ \e{4}$ & 1.33$ \e{4}$ & 5.21$ \e{3}$ & 8.34$ \e{4}$  \\
\hline
$\sigma_{g g}$ (8 TeV) [fb] & 6.79$ \e{3}$ & 6.79$ \e{3}$ & 7.00$ \e{3}$ & 2.84$ \e{3}$ & 1.11$ \e{3}$ & 1.79$ \e{4}$ \\ 
 \hline
 \hline
\end{tabular}
\caption{BPs for $Q$-only cases with no invisible decay width. Cross-sections are calculated in various diboson channels either by keeping $\Gamma_X=5$ GeV fixed (BP-1, 2, 3) or $\sigma_{\gamma \gamma}=4.8$ fb fixed (BP-4, 5, 6). We set $M_Q = 1$ TeV for all BPs. Since $\Gamma_X=5$ GeV BPs are already constrained by dijet bounds, we do not show $\Gamma_X=40$ GeV BPs in the table since dijet bounds will be even worse for them.}
\label{tab:Q_BP_no_inv}
\end{table}

It is evident from Table~\ref{tab:Q_BP_no_inv} that all BPs that fits $\Gamma_X = 5$ GeV are ruled out by CMS dijet constraint. In addition for BP-1 and BP-2 the corresponding $\sigma_{\gamma\gamma}$ are too high considering the range described by Ref.~\cite{Falkowski}. However, BP-3 provides an acceptable value of $\sigma_{\gamma\gamma}$. We did not show any BP Table~\ref{tab:Q_BP_no_inv} corresponding to $\Gamma_X = 40$ GeV since the dijet bounds are even worse for them. For $\sigma_{\gamma\gamma} = 4.8$fb cases, BP-4 and BP-5 are ruled out by the dijet bound but BP-5 survives. However, for BP-5 the total decay width of $X$ is too small (0.79 GeV), but within the $2\sigma$ range described by Ref.~\cite{Falkowski}. One might notice that for the same $\sigma_{\gamma \gamma}$, we need higher $\kappa_{gg}$ for the doublet compared to the u-type singlet resulting in higher $\sigma_{gg}$ for the doublet. This is due to the presence of moderately large $\Gamma_{WW}$ in the doublet case, which reduces the BR$(X \rightarrow \gamma \gamma)$ and requite large $\kappa_{gg}$ to achieve the desired $\sigma_{\gamma \gamma}$ value.

 This indicates that extra $L$ species may be required to increase $BR(X\rightarrow \gamma\gamma)$ to simultaneously satisfy  $\sigma_{\gamma\gamma} \sim \mathcal{O} (1-15)$ fb and $\Gamma_X \sim \mathcal{O} (5-40)$ GeV. Given the early stage of the diphoton resonance measurement, $Q$-only scenarios can be allowed for a smaller $\Gamma_X$, or very large $Q$-hypercharges in other models.

\subsection{Case II: $L \gg Q$ scenarios}

\begin{figure}[!t]
\includegraphics[scale=0.25]{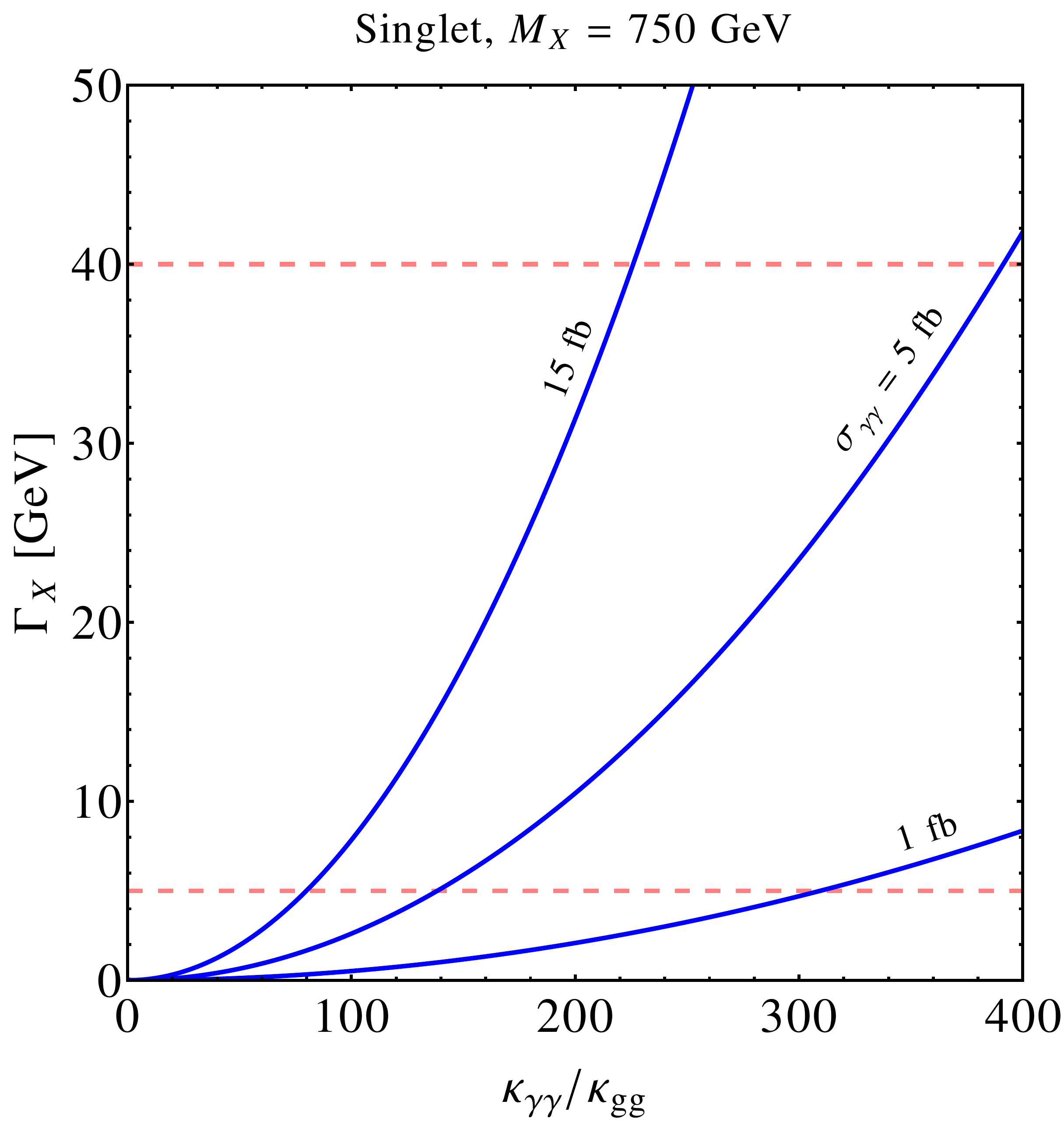}
\includegraphics[scale=0.25]{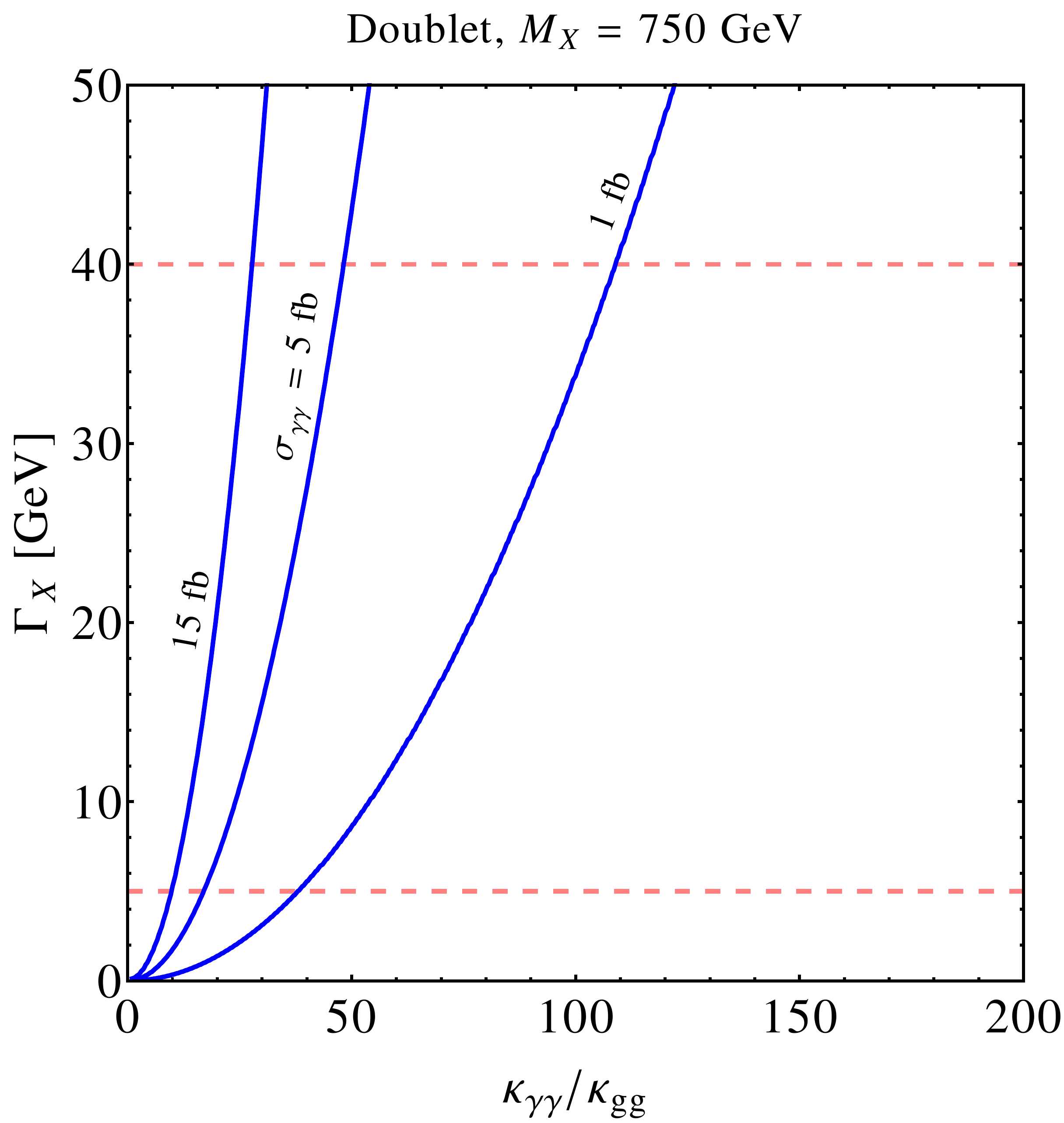}
\caption{Diphoton cross-section $\sigma_{\gamma \gamma}$ contours for different values of total decay width and the relative strength of couplings of $X$ to photons and gluons. The left panel shows the case that $L, Q$ are isosinglets and the right panel shows the isodoublet case. The pink dashed lines corresponds to the $\Gamma_X = 5$ GeV (lower) and 40 GeV (upper). }
\label{fig:sigma_kappa}
\end{figure}

When a large $L$ contribution dominates $\kappa_{\gamma\gamma}$ and $\kappa_{\gamma\gamma}\gg \kappa_{gg}$, $\kappa_{\gamma\gamma}$ also disappears from $\sigma_{\gamma\gamma}$, and we found
\be 
\kappa_{gg} =\left\{
\begin{array}{l}
5.8 \times 10^{-6} \hspace{0.1 cm} \text{GeV}^{-1}, \hspace{1cm} \text{ doublet}\\
2.1 \times 10^{-6} \hspace{0.1 cm} \text{GeV}^{-1}, \hspace{1cm} \text{isosinglet}\\
\end{array}
\right.
\ee
for $\sigma_{\gamma\gamma}=4.8$ fb. The $\Gamma_X$ can be greatly enhanced in this scenario by tuning $N_L \lambda_L$ without impacting $\sigma_{\gamma\gamma}$. In Fig.~\ref{fig:sigma_kappa} we show the  $\sigma_{\gamma \gamma}$ contours for different values of $\gamma_X$ and the relative strength of couplings of $X$ to photons and gluons. Evidently from Fig.~\ref{fig:sigma_kappa}, to simultaneously satisfy  $\sigma_{\gamma\gamma} \sim \mathcal{O} (1-15)$ fb and $\Gamma_X \sim \mathcal{O} (5-40)$ GeV we need large $\kappa_{\gamma \gamma}$ compared to $\kappa_{gg}$ for both singlet and doublet cases. 

We have noticed in the $Q$-only case that a loop-induced decay width is often small. With only $Q$s in the loop, both $\sigma_{\gamma\gamma}$ and $\Gamma_X$ are determined by $\kappa_{gg}$ and a very narrow width $\Gamma_X < 1$ GeV is expected to evade the CMS dijet bounds. Additional $L$ species would then be handy prediction if it is necessary to bring up $\kappa_{\gamma\gamma}$ and $\Gamma_X$ as illustrated in Fig.~\ref{fig:sigma_kappa}, and also suppresses $\sigma_{gg}$ at the same time. For a $\Gamma_X \sim 10^{-2} M_X$, we generally need $\kappa_{\gamma\gamma}\sim 10^2 \kappa_{gg}$. Since  $\kappa_{\gamma\gamma}\propto N_L \lambda_L$, another prediction comes in that a significant number of $L$ species must be present ($N_L \gg 1$).

In Table~\ref{tab:L_BP_no_inv} we again present two sets of BPs for doublet, u-type isosinglet cases~\footnote{d-type isosinglet cases will be similar to u-type with only $N_L \lambda_L$ may vary.}. However, as opposed to Table~\ref{tab:Q_BP_no_inv}, in this case our BPs belongs to $\Gamma_X = 5$ GeV (BP-7, 8) and $\Gamma_X=40$ GeV (BP-9, 10) respectively. Due to extra freedom available to us in $L\gg Q$ scenarios, we can easily fit $\Gamma_X=40$ GeV width with $\sigma_{gg}$ being very small. Hence, we do not show any BPs,  which satisfies $\sigma_{\gamma \gamma} = 4.8$ fb, separately since they will be very similar to $\Gamma_X=40$ GeV cases. Also note that from our discussion earlier in the section that for $\Gamma_X=40$ GeV the best-fit is obtained for $M_X=730$ GeV by Ref.~\cite{Falkowski}. So we fix $M_X=730$ GeV for $\Gamma_X=40$ GeV BPs, while still using $M_X=750$ GeV for $\Gamma_X=5$ GeV BPs for the rest of the paper with best-fit $\sigma_{\gamma \gamma}$ values 2.4 fb and 6 fb respectively (unless otherwise stated). Along with fixing $M_Q=1$ TeV, we use $M_L=400$ GeV for these calculations.

\begin{table}[!t]
\setlength\tabcolsep{5pt}
\begin{tabular}{|c|c c |c c |}
\hline\hline
 Type & doublet & u-type singlet  & doublet & u-type singlet \\
 \hline
 $L \gg Q$     & BP-7 & BP-8 & BP-9 & BP-10  \\
 \hline     
$M_X$ [GeV] & \multicolumn{2}{c|}{750} & \multicolumn{2}{c|}{730}  \\
$\Gamma_X$ [GeV] & \multicolumn{2}{c|}{5} & \multicolumn{2}{c|}{40} \\
$N_Q \lambda_Q$ & 0.17 & 0.12 & 0.25 & 0.18  \\
$N_L \lambda_L$ &  37   &  106  &  113 &  322    \\
\hline
$\kappa_{\gamma \gamma}$ [GeV$^{-1}$]   & 1.05$ \e{-4}$ & 2.99$ \e{-4}$ & 3.10$ \e{-4}$ & 8.84$ \e{-4}$  \\
$\kappa_{gg}$ [GeV$^{-1}$] & 4.31$ \e{-6}$ & 1.50$ \e{-6}$ & 6.52$ \e{-6}$ & 2.28$ \e{-6}$ \\
$\kappa_{WW}$ [GeV$^{-1}$] & 4.57$ \e{-4}$ & 0 & 1.34$ \e{-3}$ & 0  \\
\hline
$\sigma_{\gamma \gamma}$ [fb] & \multicolumn{2}{c|}{2.40} & \multicolumn{2}{c|}{6.00}  \\
$\sigma_{Z Z}$ [fb] & 7.28 & 0.20 & 18.1 & 0.49  \\
$\sigma_{Z \gamma}$ [fb] & 1.89 & 1.38 & 4.70 & 3.45   \\
$\sigma_{W W}$ [fb] & 21.2 & 0 & 52.6 & 0 \\
$\sigma_{g g}$ [fb] & 0.03 & 5$ \e{-4}$ &  0.02 & 3$ \e{-4}$   \\
\hline
$\sigma_{g g}$ (8 TeV) [fb] & 7$ \e{-4}$ & 1$ \e{-4}$ & 5$ \e{-3}$ & 7$ \e{-5}$  \\ 
 \hline
 \hline
\end{tabular}
\caption{BPs for $L \gg Q$ cases with no invisible decay width. Cross-sections are calculated in various diboson channels either for $\Gamma_X=5$ GeV  (BP-7, 8) $\Gamma_X=40$ GeV (BP-9, 10). We set $M_Q = 1$ TeV and $M_L=400$ GeV for all BPs.}
\label{tab:L_BP_no_inv}
\end{table}

Clearly from Table~\ref{tab:L_BP_no_inv} we can conclude that we require unreasonably high value of $N_L \lambda_L$ for all BPs. Hence we need $\mathcal{O}(100)$ copies  of vector-like leptons (except for BP-7), assuming perturbativity of $\lambda_L$. Nonetheless these BPs don't suffer from dijet bounds. When $L, Q$ loop contributions are comparable in $\kappa_{\gamma\gamma}$, i.e. $N_Q \lambda_Q \sim N_L \lambda_L$, we will again suffer from narrow-width and large dijet cross-section problems. We do not discuss this mixed case here for simplicity. However, even in that case we will need many copies of both $Q$ and $L$, comparable to numbers shown in Table~\ref{tab:Q_BP_no_inv}.

\subsection{Possible invisible decay of $X$ }

To solve the $L,Q$ multiplicity issue, it is possible to couple $X$ to complete SM singlets $N$, via for instance $X\bar{N}N$, and such $N$ can have a mass below $M_X/2$ and $X$ can decay into $\bar{N},N$ at tree level. This invisible width can solve the very-narrow width issue of the new resonance. In this light, an economical setup can be $~4$ isodoublet $Q$ and $L$ species to give the correct $\sigma_{\gamma\gamma}$ with a narrow width, and the invisible $X$ decays to accommodate for the measured $X$ width.
However monojet can have severe constraint if $N$ is a dark matter particle. If however N decays to the DM particle with the emission of a virtual Z with a small mass gap $\sim$ 30-40 GeV, we can evade the monojet bound due to these leptons and jets. Since mass gap is not large and N's are produced back to back the amount of missing energy is small and there exists almost no constraint on these small gap situations.

We present few more BPs in Table~\ref{tab:Q_BP_inv} and \ref{tab:L_B_inv}, taking into account large invisible width, for both $Q$-only and $L \gg Q$ scenarios with $\Gamma_X=5$ and 40 GeV. Now with the addition of extra parameter $\Gamma_{\text{inv}}$ we can find BP with $\Gamma_X=40$ GeV for $Q$-only cases also. Therefore, similar to Table~\ref{tab:L_BP_no_inv}, we don't show BPs with $\sigma_{\gamma \gamma} = 4.8$ fb cases. 

We see from Table~\ref{tab:Q_BP_inv} that with the introduction of large invisible width we reduce $N_Q \lambda_Q$ by a factor of $\sim 2$ for $\Gamma_X=5$ GeV. In contrast, for $\Gamma_X=40$ GeV we still need $N_Q \lambda_Q \sim \mathcal{O}(10)$. Most BPs in Table~\ref{tab:Q_BP_inv} evade the dijet bound of 2 pb at 8 TeV. We note that for BP-13 and BP-15 we tune the parameters to evade the aforementioned bound, which in turn reduces the $\sigma_{\gamma \gamma}$. Nevertheless, the  values presented in Table~\ref{tab:Q_BP_inv} are within $95 \%$ CL range prescribed by Ref.~\cite{Falkowski}. However, for BP-16 (d-type quark singlet with $\Gamma_X = 40$ GeV) even for the lowest allowed $\sigma_{\gamma \gamma}$ value allowed, we could not avoid the dijet bound.   

\begin{table}[h]
\setlength\tabcolsep{5pt}
\begin{tabular}{|c|c c c|c c c|}
\hline\hline
 Type & doublet & u-type singlet & d-type singlet & doublet & u-type singlet & d-type singlet \\
 \hline
  $Q$-only with invisible decays   & BP-11 & BP-12 & BP-13 & BP-14 & BP-15 & BP-16 \\
 \hline     
$M_X$ [GeV] & \multicolumn{3}{c|}{750} & \multicolumn{3}{c|}{730}  \\
$\Gamma_X$ [GeV] & \multicolumn{3}{c|}{5} & \multicolumn{3}{c|}{40} \\
$N_Q \lambda_Q$ & 3.47 & 5.49 & 7.5 & 6.27 & 11.6 & 17.7 \\

$\Gamma_{\text{inv}}$ [GeV] & 2.76 & 3.62 & 2.42 & 33.3 & 34.3 & 26.8 \\
\hline
$\kappa_{\gamma \gamma}$ [GeV$^{-1}$] & 4.96$ \e{-6}$ & 6.82$ \e{-6}$ & 2.15$ \e{-6}$ & 8.95$ \e{-6}$ & 1.33$ \e{-5}$ & 5.06$ \e{-6}$   \\
$\kappa_{gg}$ [GeV$^{-1}$] &  9.07$ \e{-5}$ & 7.17$ \e{-5}$ & 9.80$ \e{-5}$ & 1.63$ \e{-4}$ & 1.52$ \e{-4}$ & 2.31$ \e{-4}$ \\
$\kappa_{WW}$ [GeV$^{-1}$] & 3.87$ \e{-5}$ & 0 & 0 & 6.97$ \e{-5}$ & 0 & 0 \\
\hline
$\sigma_{\gamma \gamma}$ [fb] & 2.40 & 2.41 & 0.52 & 3.17 & 6.00 & 2.02  \\
$\sigma_{Z Z}$ [fb] & 20.0 & 0.20 & 0.04 & 26.4 & 0.49 & 0.17 \\
$\sigma_{Z \gamma}$ [fb] & 11.6 & 1.38 & 0.30 & 15.2 &  3.44 & 1.16  \\
$\sigma_{W W}$ [fb] & 67.9 & 0 & 0 & 89.4 & 0  &  0 \\
$\sigma_{g g}$ [fb] & 6.41$ \e{3}$ & 3.51$ \e{3}$ & 8.75$ \e{3}$ & 8.47$ \e{3}$ & 6.25$ \e{3}$ & 3.37$ \e{4}$  \\
$\sigma_{\text{inv}}$ [fb] & 8.01$ \e{3}$ & 6.57$ \e{3}$ & 8.21$ \e{3}$ & 4.26$ \e{4}$ & 3.77$ \e{4}$ & 6.83$ \e{4}$  \\
\hline
$\sigma_{g g}$ (8 TeV) [fb] & 1.37$ \e{3}$ & 535 & 1.86$ \e{3}$ & 1.84$ \e{3}$ & 1.36$ \e{3}$ & 7.31$ \e{3}$ \\ 
 \hline
 \hline
\end{tabular}
\caption{BPs for $Q$-only cases with large invisible decay width. Cross-sections are calculated in various diboson channels  by keeping $\Gamma_X$ fixed at 5 GeV (BP-11, 12, 13) and  40 GeV (BP-14, 15, 16). We set $M_Q = 1$ TeV for all BPs.}
\label{tab:Q_BP_inv}
\end{table}

\begin{table}[h]
\setlength\tabcolsep{5pt}
\begin{tabular}{|c|c c |c c |}
\hline\hline
 Type & doublet & u-type singlet  & doublet & u-type singlet \\
 \hline
 $L \gg Q$ with invisible decays & BP-17 & BP-18 & BP-19 & BP-20  \\
 \hline     
$M_X$ [GeV] & \multicolumn{2}{c|}{750} & \multicolumn{2}{c|}{730}  \\
$\Gamma_X$ [GeV] & \multicolumn{2}{c|}{5} & \multicolumn{2}{c|}{40} \\
$N_Q \lambda_Q$ & 2.02 & 2.92 & 4.30 & 6.30  \\
$N_L \lambda_L$ & 2.00 & 3.00 & 4.31 & 6.30   \\
$\Gamma_{\text{inv}}$ [GeV] & 4.2 & 4.6 & 36.7 & 38.3 \\
\hline
$\kappa_{\gamma \gamma}$ [GeV$^{-1}$]   & 8.53$ \e{-6}$ & 1.18$ \e{-5}$ & 1.80$ \e{-5}$ & 2.45$ \e{-5}$  \\
$\kappa_{gg}$ [GeV$^{-1}$] & 5.28$ \e{-5}$ & 3.82$ \e{-5}$ & 1.12$ \e{-4}$ & 8.22$ \e{-5}$ \\
$\kappa_{WW}$ [GeV$^{-1}$] & 4.69$ \e{-5}$ & 0 & 9.90$ \e{-5}$ & 0  \\
\hline
$\sigma_{\gamma \gamma}$ [fb] & \multicolumn{2}{c|}{2.40} & \multicolumn{2}{c|}{6.00}  \\
$\sigma_{Z Z}$ [fb] & 10.8 & 0.20 & 27.1 & 0.49  \\
$\sigma_{Z \gamma}$ [fb] & 4.22 & 1.38 & 10.6 & 3.44   \\
$\sigma_{W W}$ [fb] & 33.9 & 0 & 84.7 & 0 \\
$\sigma_{g g}$ [fb] & 736 & 201 &  1.87$ \e{3}$ & 540 \\
$\sigma_{\text{inv}}$ [fb] & 4.13$ \e{3}$ & 2.37$ \e{3}$ & 2.21$ \e{4}$ & 1.24$ \e{4}$ \\
\hline
$\sigma_{g g}$ (8 TeV) [fb] & 157 & 43 & 407 & 117  \\ 
 \hline
 \hline
\end{tabular}
\caption{BPs for $L \gg Q$ cases with large invisible decay width. Cross-sections are calculated in various diboson channels either for $\Gamma_X=5$ GeV  (BP-17, 18) $\Gamma_X=40$ GeV (BP-19, 20). We set $M_Q = 1$ TeV and $M_L=400$ GeV for all BPs.}
\label{tab:L_B_inv}
\end{table}

\begin{figure}[!t]
\includegraphics[scale=0.155]{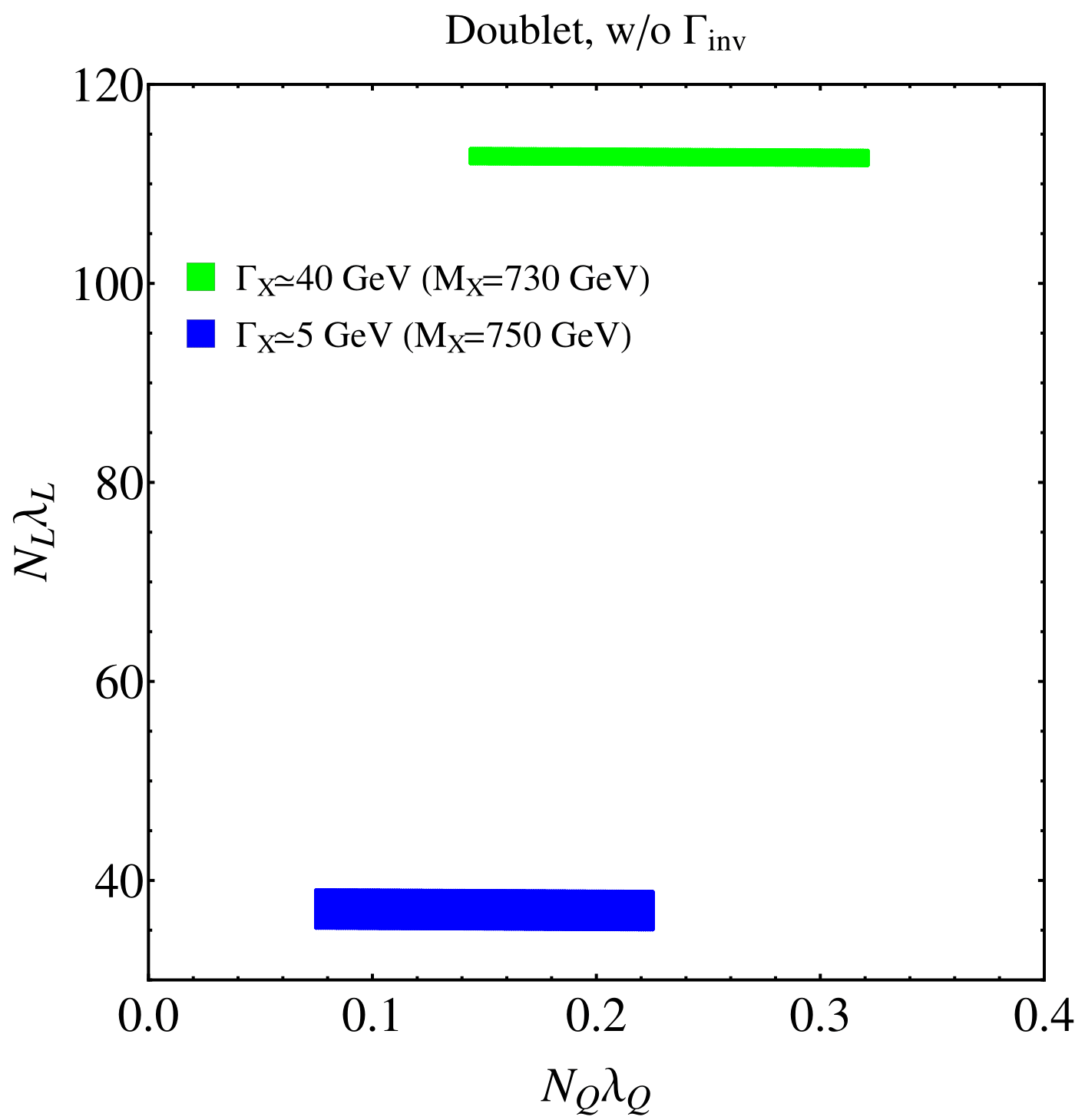}
\includegraphics[scale=0.15]{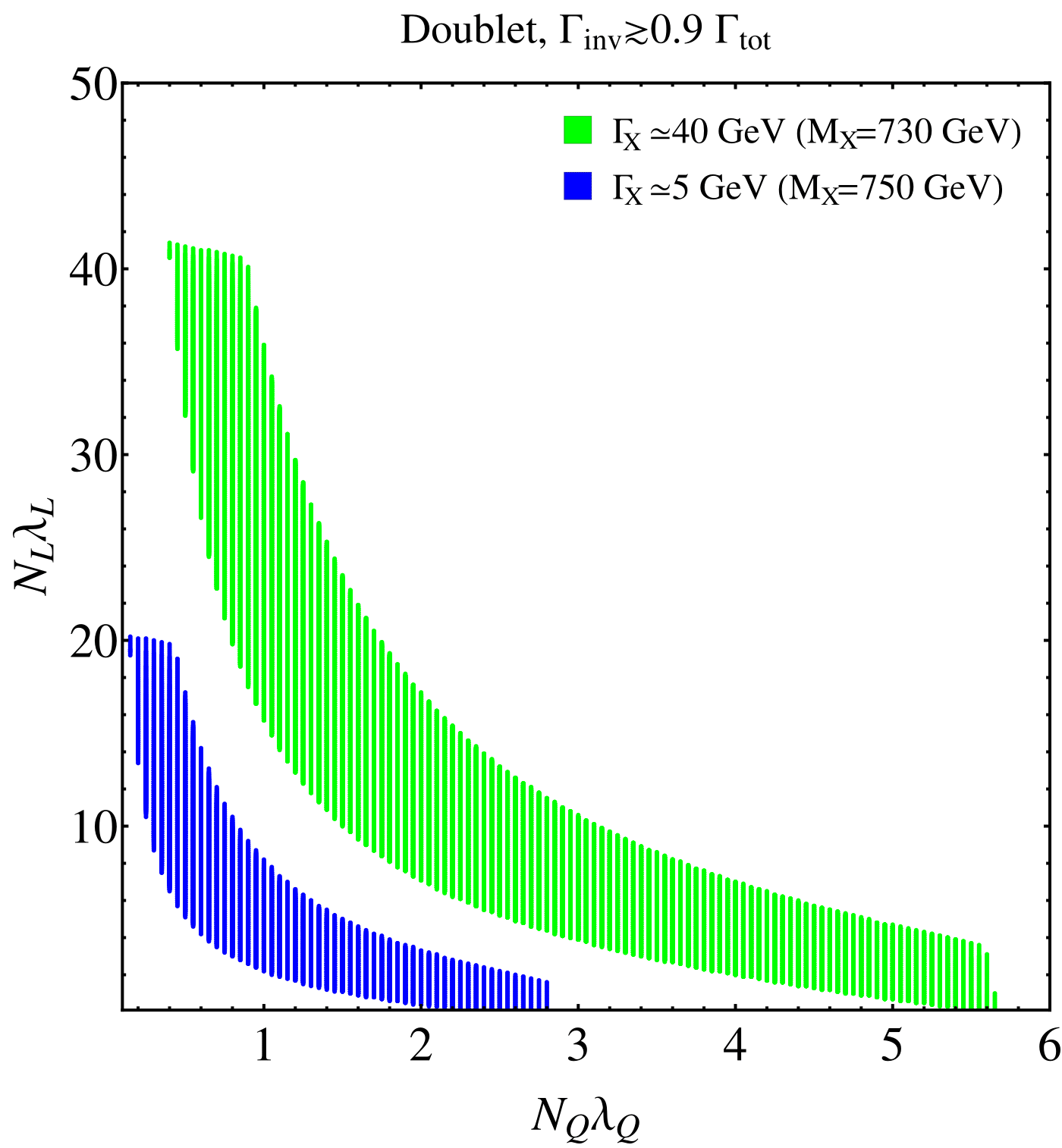}
\caption{Doublet $N_L\lambda_L$, $N_Q\lambda_Q$ ranges for different values of the resonance width. The left panel assume no invisible decays and the right panel assume a dominant invisible width of $X$.}
\label{fig:doubletnumbers}
\end{figure}

\begin{figure}[!t]
\includegraphics[scale=0.155]{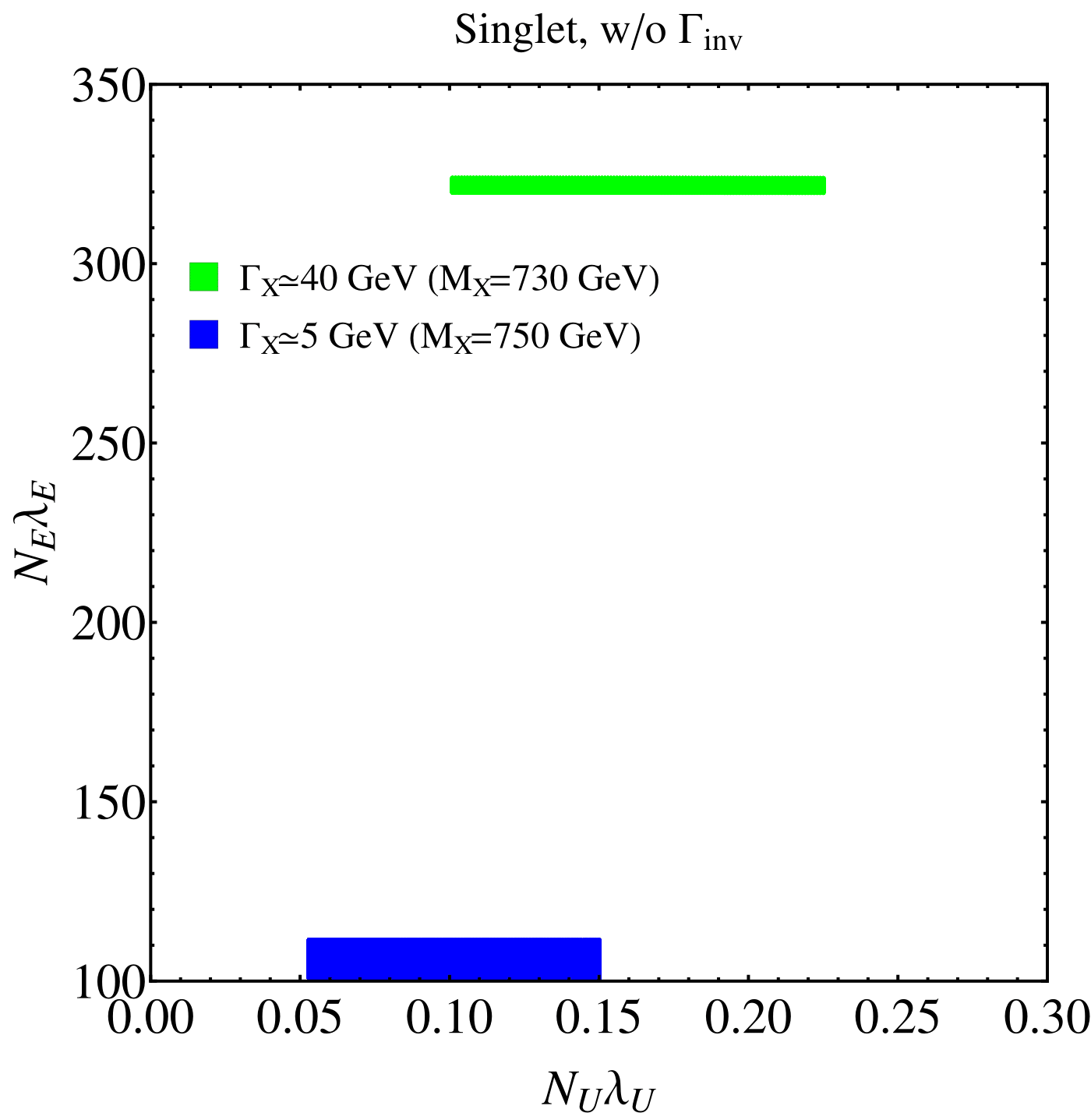}
\includegraphics[scale=0.15]{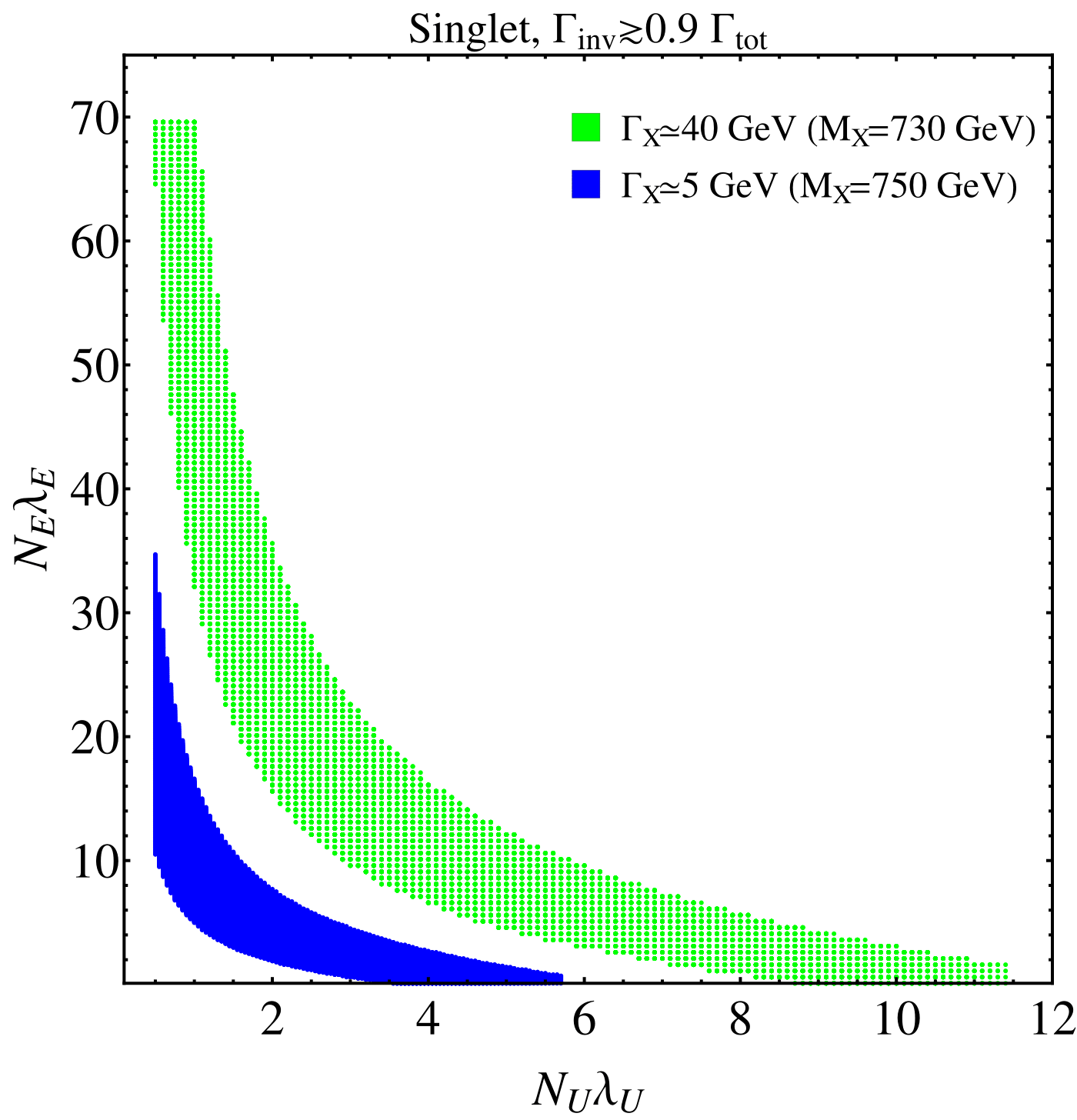}
\caption{Singlet $N_L\lambda_L$, $N_Q\lambda_Q$ ranges for different values of the resonance width. The left panel assume no invisible decays and the right panel assume a dominant invisible width of $X$.}
\label{fig:singletnumbers}
\end{figure}

In Table~\ref{tab:L_B_inv} we tabulate BPs with the most economical choice of both $N_Q \lambda_Q$ and $N_L \lambda_L$. The inclusion of  large invisible width reduces the multiplicity of $L$ to a very reasonable level ($\sim 2-6$). These BPs again evade the dijet bound. Finally, in Figs.~\ref{fig:doubletnumbers} and \ref{fig:singletnumbers}, we show the combinations of $N_Q\lambda_Q$ and $N_L\lambda_L$ for the total decay width =5 and 40 GeV with and without $X\rightarrow N\bar{N}$ which we call $\Gamma_{\text{inv}}$. We see again that the presence of the invisible width allows us to fit the width with smaller values of $N_Q\lambda_Q$ and $N_L\lambda_L$ both for doublet and singlet fields.


\section{Associated collider tests}
\label{sect:associated}

If the 750 GeV diphoton excess persists, a few associated signal would be expected from the mixing between the SM gauge fields:

\ \ (i) $X\rightarrow ZZ$, $Z\gamma$ decays, leading to $4l$, $2l+\met/\gamma$, $\gamma +\met$ channels

\ \ (ii) An $X\rightarrow W^+W^-$ final state if $L$ and/or $Q$ is an isospin doublet.
  
The expected signal cross-section is readily given by $\sigma_{VV} = \sigma_{\gamma\gamma}\frac{BR(X\rightarrow VV)}{BR(X\rightarrow \gamma\gamma)}$, multiplied by further decay branchings of the SM vector bosons.
Here we list the leading predicted signals in Table~\ref{tab:associated_channels} for all BPs considered in the previous section that survive the CMS dijet bound. 
The presence of such associated decays should serve as good test of the SM gauge mixing, if the 750 GeV resonance is established in the future data. Alternatively, these $ZZ,Z\gamma$ channels can help confirm/rule out new physics scenarios, e.g. our weakly charged vector-like heavy fermion hypothesis.



\begin{table}[!t]
\begin{tabular}{|c|c|c|c|c|c|}
\hline \hline
Channel & $ZZ\rightarrow 4l$ &  $ZZ\rightarrow 2l+\met$ & $Z\gamma\rightarrow \gamma +2l$ & $Z\gamma\rightarrow \gamma +\met$ & $WW\rightarrow{e\mu +\met}$ \\
BR & $0.45\%$ & $2.7\%$ & $6.7\%$ & $20 \%$ & $2.3\%$ \\
\hline
note & two pairs of $M_{ll}=M_Z$ & $M_{ll}=M_Z$ & mono-photon,$M_{ll}=M_Z$ & mono-photon with large $\met$ &  different $l$ flavor  \\
\hline
BP-5 & 0.002 & 0.01 & 0.19 & 0.57 & 0 \\
BP-7 & 0.03 & 0.20 & 0.13 & 0.38 & 0.49 \\
BP-8 & 0.0009 & 0.005 & 0.09 & 0.28 & 0 \\
BP-9 & 0.08 & 0.49 & 0.31 & 0.94 & 1.21 \\
BP-10 & 0.002 & 0.01 & 0.23 & 0.69 & 0 \\
BP-11 & 0.09 & 0.54 & 0.78 & 2.32 & 1.56 \\
BP-12 & 0.0009 & 0.005 & 0.09 & 0.28 & 0 \\
BP-13 & 0.0002 & 0.001 & 0.02 & 0.06 & 0 \\
BP-14 & 0.12 & 0.71 & 1.02 & 3.04 & 2.06 \\
BP-15 & 0.002 & 0.01 & 0.23 & 0.69 & 0 \\
BP-17 & 0.05 & 0.29 & 0.28 & 0.84 & 0.78 \\
BP-18 & 0.0009 & 0.005 & 0.09 & 0.28 & 0 \\
BP-19 & 0.12 & 0.73 & 0.71 & 2.12 & 1.95 \\
BP-20 & 0.002 & 0.01 & 0.23 & 0.69 & 0\\

\hline \hline
\end{tabular}
\caption{A few leading test channels arising from associated $X\rightarrow VV$ decays, for BPs which survive the CMS dijet bound. Here $l=e, \, \mu$ refers only to the first two generation leptons, and $\met$ arises from neutrinos. All the cross-sections are in fb.}
\label{tab:associated_channels}
\end{table}

The 4$l$ and the mono-photon +$\met$ channels are probably the most imminent tests of the associate $X\rightarrow ZZ,Z\gamma$ decays. In current data, CMS~\cite{Chatrchyan:2012rva} constrains a $ZZ$ resonance at 750 GeV to be less than 0.12 pb, The associated mono-photon signal in the $Q$-only doublet cases are close to be constrained but within the existing CMS ~\cite{Khachatryan:2014rwa} limits. Future updates from 13 TeV runs would strongly constrain or confirm these associated signals.

\section{Theoretical discussions }
\label{sect:theory_discussion}

There are many motivation to look beyond the SM physics. Namely neutrino
masses and mixings, dark matter,
gauge coupling unification and the SM Higgs mass vacuum stability.
Recently, the CMS and ATLAS Collaborations
announced excess in the distribution of events containing two photons
peaked at 750 GeV or so can interpret
as new motivation of physics beyond the SM. Our goal in this paper is to
connect all above mentioned motivation
of new physics to each other.

\begin{figure}[!t]
\centering
\includegraphics[scale=0.9]{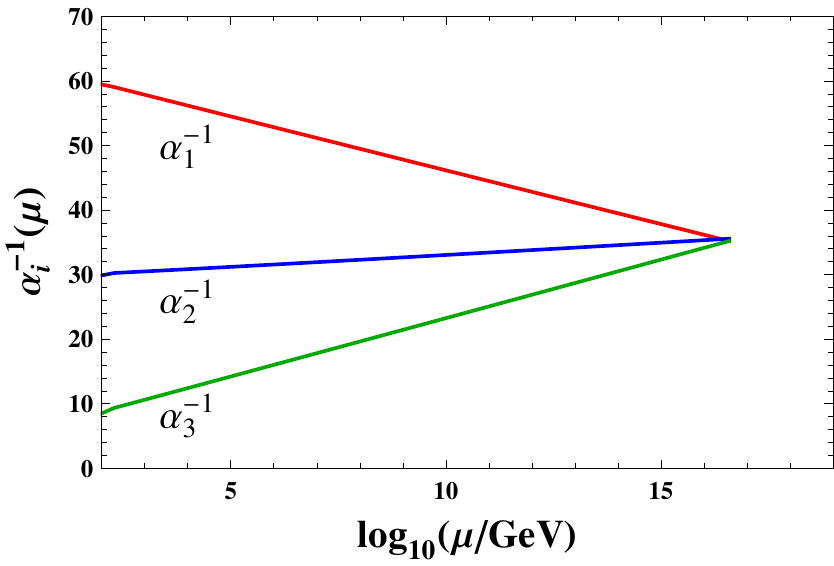}
\caption{Gauge coupling unification at two loop involving fermions:
   $Q(3,2,1/6) +  \overline{Q}(\overline{3},2,-1/6)$,
 Scalars:~ ${D^c_s }(3,1,-1/3) + {U^c_s}(3, 1,2/3)$. Unification happens at $ 7\times 10^{15}$ GeV with $\alpha^{-1}=36$.
\label{figure2}
}
\end{figure}

It was shown some time ago that demanding gauge coupling unification to
be consistence proton decay constraint
can lied simplest and most minimal extension of the SM as follows
\begin{eqnarray}
Q_f\left(3, 2, \frac{1}{6}\right) +\overline{Q}_f\left(\overline{3}, 2,
-\frac{1}{6}\right)+ D^c_s\left(3, 1, -\frac{1}{3}\right)+
{U^c_s}\left( 3, 1, \frac{2}{3}\right) + L^c_s \left(1, 2, \frac{1}{2}\right),
\label{nf1}
\end{eqnarray}

where subscripts $s$ and $f$ are for scalars and fermions of new
particles. We would like to emphasize that
this model can be realized in the orbifold
Grand Unified Theories (GUT). The result performing two loop RGE evaluation is presented in Fig.~\ref{figure2}.

\begin{figure}[!t]
\centering
\includegraphics[scale=0.7]{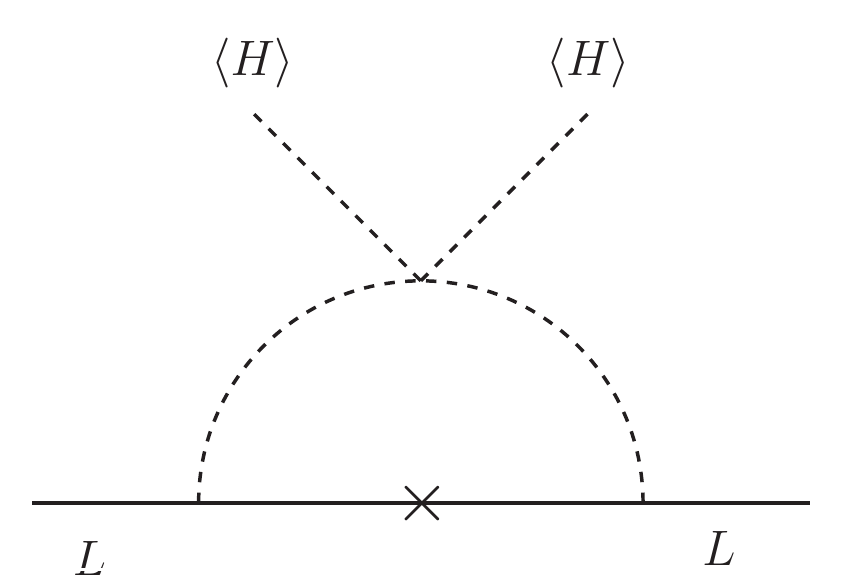}
\caption{One loop diagram for generating neutrino masses.
\label{figure3}
}
\end{figure}

As shown in ref. \cite{Babu:2011vb} the neutrino masses and mixings can
be generated radiatively using the Lagrangian
\begin{eqnarray}\mathcal{L}\supset M_Q Q\overline{Q} + M^2_{D^c} |D_s^c|^2 + M^2_{U^c}|U_s^{2}|^2 + \lambda_1 Q L D_s^c + \Lambda_2 \overline Q L {{U_s^{c\ast}}} + \lambda D_s^c {U_s^{c\ast}} H^2
\label{nf2}
 \end{eqnarray}
The loop involves 
colored particle presented in Eq.~(\ref{nf1}) (shown in 
Fig.~\ref{figure3}) and the expression for neutrino mass:

\begin{eqnarray}
\mathcal{M}_\nu \simeq \frac{\lambda_1\lambda_2}{16\pi^2} \frac{\lambda 
\langle H \rangle ^2}{M^2_{D^c}-M^2_{U^c}} \left( \frac{M_{D^c}^2 
M_Q}{M^2_Q-M_{D^c}^2}\log\frac{M^2_Q}{M_{D^c}^2} - \frac{M_{U^c}^2 
M_Q}{M^2_Q-M_{U^c}^2}\log\frac{M^2_Q}{M_{U^c}^2}\right)
\end{eqnarray}

On the other hand, having in the spectrum stable scalar doublet
can be interpreted as inert doublet model for dark matter
\cite{Ma:2006km}. The charge neutral component of the doublet can be
lighter than the charged component due to radiative
corrections~\cite{Garcia-Cely:2015khw}. Also it is very interesting
to note that having this (See Eq. (\ref{nf1})) new particle in the
spectrum can solve the Higgs vacuum stability
problem \cite{Gogoladze:2010in}. It is very interesting to note that all
this new physics motivation
can lead to the prediction to the diphoton excess which we hope will be
confirmed in near future.

We would like to point out that the generic vector-like
particles do not need to form complete $SU(5)$
or $SO(10)$ representations in Grand Unified Theories (GUTs) from the
orbifold constructions~\cite{kawa, GAFF, LHYN},
intersecting D-brane model building on Type II
orientifolds~\cite{Blumenhagen:2005mu, Cvetic:2002pj, Chen:2006ip},

M-theory on $S^1/Z_2$ with Calabi-Yau
compactifications~\cite{Braun:2005ux, Bouchard:2005ag}, and
F-theory with $U(1)$ fluxes~\cite{Donagi:2008ca, Beasley:2008dc, Beasley:2008kw, Donagi:2008kj} (See Ref.~\cite{Li:2010hi} and references therein.)

The generic vector-like particles from orbifold GUTs and F-theory GUTs
have been studied previously in 
Refs.~\cite{Jiang:2009za, Li:2009cy, Li:2010hi}. From Ref.~\cite{Li:2010hi}, 
we found that in the orbifold GUTs, we cannot realize such set of vector-like
particles and scalars since the scalar $U_s^c$ cannot be obtained.
Interestingly, in the F-theory $SU(5)$ models, we can indeed realize
such set of vector-like particles and scalars. For details, please
see Table IV in Ref.~\cite{Li:2010hi}. Moreover, without additional vector-like
particles or scalars, the GUT scale is still around $7\times 10^{15}$ GeV,
and the GUT gauge coupling $\alpha$ is about 1/36. Thus, the
proton lifetime via dimension-6 operators will be within the reach
of the future Super-Kamiokande and Hyper-Kamiokande experiments.

\section{Conclusion}
\label{sect:conclusion}

We proposed a new model from orbifold GUTs/string models to explain the recent diphoton resonance 
at the LHC by introducing new scalars and vector-like fermions. We showed that 
it is possible to explain the diphoton resonance, and such
diphoton resonance explanation does not conflict with any other bound. Interestingly, 
the upcoming results can constrain some of these explanations. We investigated the number of copies of new particles a to explain the excess. We noticed that the new fermions 
and scalars are also helpful to provide us grand unification of gauge couplings. 
Further, the new fields also generate neutrino masses and the non-colored doublet provide 
the dark matter candidate. In addition the vector like fields also provides us 
a stability of the electroweak vacuum since the SM gauge couplings become strong at high scale
via RGE running. We showed the constraints on the new couplings and numbers of new multiplets.

Note added: While we are completing the draft, we noticed a few papers~\cite{Harigaya:2015ezk,Mambrini:2015wyu,Backovic:2015fnp,Angelescu:2015uiz,Nakai:2015ptz,Knapen:2015dap,Buttazzo:2015txu,Pilaftsis:2015ycr,Franceschini:2015kwy,DiChiara:2015vdm,McDermott:2015sck,Kobakhidze:2015ldh,Ahmed:2015uqt} appeared in the arXiv on the same topic.

\bigskip
{\bf Acknowledgement}

We thank Ali Celic, Mykhailo Dalchenko and Teruki Kamon for helpful discussions. This work is supported by DOE Grant DE-FG02-13ER42020 (B.D., T.G.), Natural Science Foundation of China under grant numbers 11135003, 11275246, and 11475238 (TL) and support from the Mitchell Institute. Y.G. thanks the Mitchell Institute for Fundamental Physics and Astronomy for support. I.G. thanks the Bartol Research Institute for partial support.

\end{document}